# "Self aligned" setup for laser optical feedback imaging insensitive to parasitic optical feedback.

Olivier Jacquin, Samuel Heidmann, Eric Lacot, and Olivier Hugon,
*Laboratoire de Spectrométrie Physique, Université Joseph Fourier de Grenoble, UMR CNRS 5588, B.P. 87, 38402 Saint martin d'Hères Cedex, France -*
*Corresponding author: ojacquin@ujf-grenoble.fr*

In this paper we propose a new optical architecture for the laser optical feedback imaging (LOFI) technique which makes it possible to avoid the adverse effect of the optical parasitic backscattering introduced by all the optical interfaces located between the laser source and the studied object. This proposed setup need no specific or complex alignment, that why we can consider the proposed setup as self aligned. We describe the principle used to avoid the parasitic backscattering contributions which deteriorate dramatically amplitude and phase information contained in the LOFI images. Finally, we give successful demonstration of amplitude and phase images obtained with this self aligned setup in presence of a parasitic reflection.

## LOFI technique.

The LOFI technique is a very sensitive imaging method combining optical heterodyne interferometry with the dynamic properties of class B lasers [1]. In this method, the interference takes place into the laser, between the intracavity light and the backscattered light by the studied target. The backscattered light is frequency shifted in order to create an intracavity optical beating. The laser output power is then modulated at the shift frequency $\Omega$. If the shift frequency $\Omega$ is resonant with laser relaxation frequency $\Omega_R$, a great optical amplification of the optical beating contrast can be obtained. The detection of this modulation with a lock-in amplifier makes it possible to realize simultaneously amplitude (*i.e.* reflectivity) images and phase (*i.e.* profilometry) images of a non cooperative targets [2]. For example, with a Nd:YAG microchip laser, the amplification is of the order of $10^6$, which makes it possible to easily measure reflectivity as low as $10^{-13}$ with a laser output power of a few milliwatts and with a bandwidth detection of 1 KHz [3].

In the LOFI technique, the laser and the target are conjugated via the optics of the system and the backscattered photons come back into the laser cavity according to the reverse path principle. The system is then self-aligned since laser is both used as source and detector (fig.1). Consequently, the optical system needs no complex alignment, which is another great advantage of the LOFI technique.

Figure 1 shows a description of the LOFI experimental setup. The laser is a cw $Nd^{3+}$:YAG microchip lasing at the wavelength $\lambda$=1064 nm with an output power of 1 mW and a relaxation frequency $\Omega_R=700$ kHz. A two-axis galvanometric mirror scanner makes it possible to move the laser beam on the surface of the studied target to build point by point an image. The frequency shift is obtained with two acousto-optic deflectors (AOD) operating respectively at 81.5MHz (order +1) and 81.5MHz-$\Omega/2$ (order -1). The frequency shift is equal to $\Omega/2$ when the light passes through the shifter. After a round trip, the total frequency shift of the reinjected light into the laser is thus equal to $\Omega$.

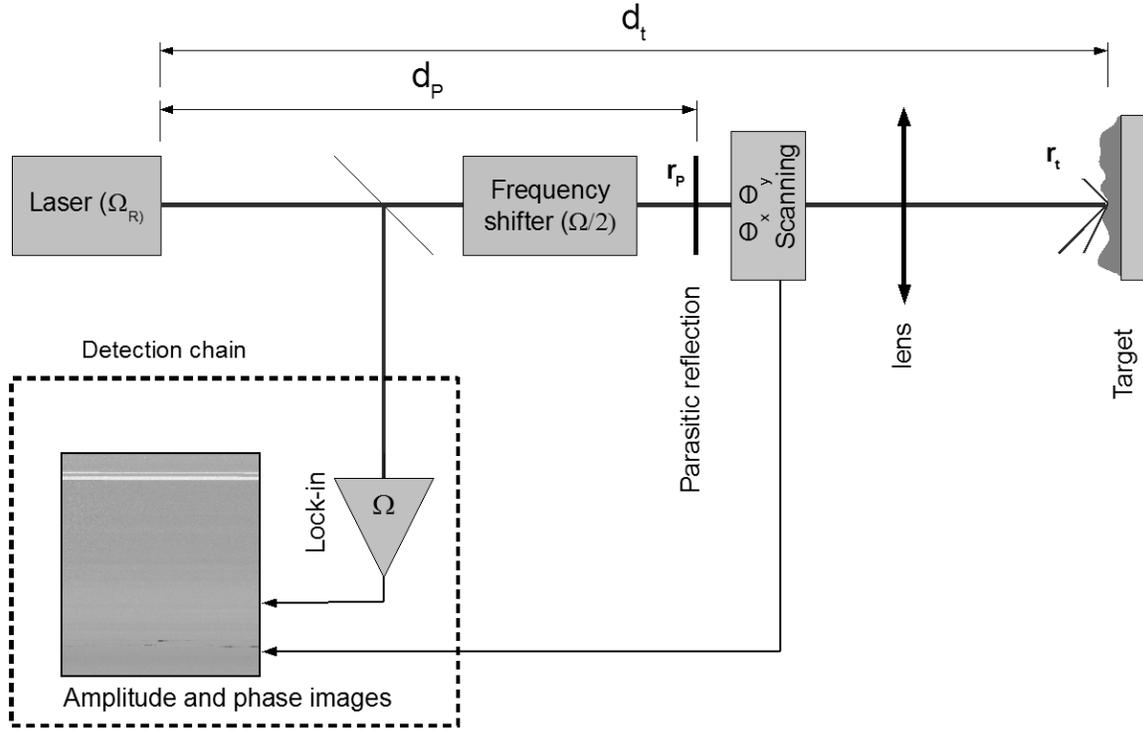

*Figure 1: description of the classical LOFI experiment.*

## Parasitic optical feedback

The LOFI method is extremely sensitive to all optical feedback. Consequently the method is sensitive to optical parasitic backscattering which is inherent in all optical systems [4][5] and which depends on the quality of the optical elements. A significant optical parasitic backscattering generated by an optical element located between the frequency shifter and the studied target limits dramatically the LOFI performances. Indeed, we intuitively understand that it may be difficult to detect reinjected target signal lower than that one reinjected by "the parasitic object". The sensitivity of the LOFI technique is then strongly limited. For a "parasitic" diffusing object with an effective reflectivity $r_P$ and for a target with an effective reflectivity $r_t$, respectively located at distances $d_P$ and $d_t$ from the laser (fig.1), the expressions of amplitude and phase extracted by the lock-in amplifier are [6]:

$$R = G_{LOFI}\sqrt{r_t^2 + r_P^2 + 2r_t r_P \cos(\phi_t - \phi_p)}P_{out} \qquad (1)$$

$$\phi = a\tan\left[\frac{r_t \sin(\phi_t) + r_P \sin(\phi_p)}{r_t \cos(\phi_t) + r_P \cos(\phi_p)}\right] \quad (2)$$

with $\phi_p = \frac{2\pi}{\lambda}2d_P$, $\phi_t = \frac{2\pi}{\lambda}2d_t$, $P_{out}$ is the output laser power and $G_{LOFI}$ is the optical amplification of the optical beating contrast [2].

The previous equations show that for a significant parasitic reflection ($r_P \approx r_t$) the amplitude ($r_t$) and phase ($\phi_t$) image of the target may be inaccessible as we have shown in [7]. To illustrate this effect, we give in Fig.2, the phase ($\phi$) and amplitude R images obtained with a controlled parasitic reflection in the experimental setup (fig.1). The parasitic optical feedback is generated by a microscope slide placed between the target and the frequency shifter. The orientation of the slide is accurately adjusted in order to obtain $r_P \approx r_t$. For the amplitude images, the studied object is a piece of metallic ruler while for the phase image it is the letter "e" etched on silicon plate (the typical dimension of the letter is about hundred micrometers). The phase images given in Fig.2 have been unwrapped using the maximum-likelihood binary-tree (MLBT) method [8].

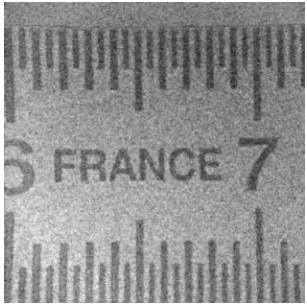

*fig.2a)*

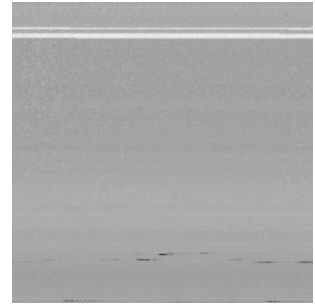

*fig.2 c)*

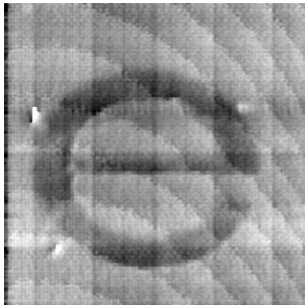

*fig.2b)*

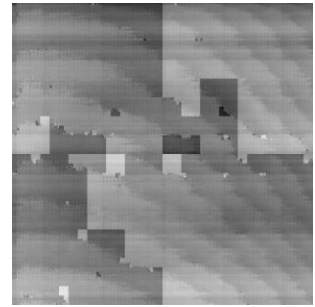

*fig.2 d)*

*figure 2: image obtained with and without a significant parasitic reflection ($r_P \approx r_t$) in the classical LOFI setup (All images have been realized at the frequency Ω). a) Amplitude image without parasitic reflection b) Phase image without parasitic reflection c)Amplitude image with parasitic reflection d) Phase image without parasitic reflection*

The figures 2a) and 2b) show respectively amplitude and phase image obtained without parasitic backscattering. We can see on the figures 2c) and 2d) the effect of parasitic reflection on these images. The phase and amplitude information is completely scrambled. Indeed, in the amplitude image (fig.2c) we no longer recognize the pattern of metallic ruler, and in the phase image (fig.2d) we no longer distinguish the profile of the etched letter. The rectangular patterns in the fig.2 are typical patterns of the MLTB method when the phase information phase cannot be unwrapped.

The use of anti-reflective (AR) coated optics [9] at the working wavelength in the optical device makes it possible to minimize this critical effect introduced by the parasitic backscattering. However, problems arise when we wish to satisfy the inequality $r_p<<r_t$ for $r_t$ reflectivity as low as $10^{-13}$. For example, if the LOFI technique is coupled with a microscope to realize biological images [10], it is difficult to find microscope objectives with AR coating at the wavelength of 1.064µm, and very expensive or not reasonable to use AR coated microscope slides. The surface of the sample may also cause an important echo that limits the possibilities of investigations under this surface. This example highlights the need to eliminate the adverse effects caused by the optical parasitic backscattering. We have proposed in a previous paper [7] an architecture that makes it possible to avoid these effects. However, this device is not self aligned which complicate considerably the implementation of the LOFI optical setup. This motivates us to develop a new architecture insensitive to parasitic reflection that allows to keep the self aligned feature and so the simplicity of the classical LOFI setup.

## Anti-reflection LOFI device

We propose a self aligned LOFI device (fig.3) allowing the detection of backscattered light by the studied target without optical parasitic backscattering contribution. The laser beam is splitted in two parts, using double refraction phenomena in an anisotropic material plate. This splitter is beam displacer (Melles griot reference: [03 PBD 001](03 PBD 001)) which split a beam of light into two mutually orthogonal, linearly polarized beams that are parallel to one another and to the axis of the input beam. In our experiment, the laser beam displacement is around 2.7mm. The laser beam polarization at the entrance of the splitter is controlled by means of a polarizer combined with a half-wavelength retardation plate. To control the return path of backscattered light from a studied target or from a parasitic reflection source, a quarter-wavelength retardation plate is placed after the frequency shifter. The control principle is explained in the next paragraph. The slow and fast axes of this plate are oriented at the 45° angle to polarizations of both orthogonally polarized laser beams. Both laser beams are finally focused in a single point of the target by a lens. The scanning is still performed by two galvanometric mirrors. The LOFI experimental setup presented in Fig.3 needs no specific or/and complex alignment, it is self aligned. Indeed, after the beam splitter, both laser beams are parallel and a priory they overlap in image plan of the focusing lens. The only critical alignment is the centering of the laser beams with the lens, in order to limits astigmatism aberration, as in the classical LOFI setup. In the setup of the fig.3 this alignment may be more difficult than in a classical LOFI setup because the beam has been split into two beams. However, it is even less difficult than the distance between both laser beams is small.

In this LOFI device, the optical path (up or down in the fig.3) taken by the backscattered light may be different for the parasitic backscattered light and for the backscattered light coming from the target. In the case of parasitic reflection which takes place outside of the overlapping point of both laser beams, the light necessarily goes back to the laser by the same path. On the other hand, the backscattered light by the target (*i.e.,* the signal) may come back toward the laser by a different path. The roundtrip path of the backscattered light is thus an optical loop. Depending on the roundtrip path (identical path for incident and backscattered light or loop path), the polarization and the frequency shift of the backscattered light are different which allows to differentiate the parasitic backscattering from the target backscattering. This filtering may be optical (via the polarization) and/or electronic (via the frequency shift) depending on the position of the parasitic backscattering source.

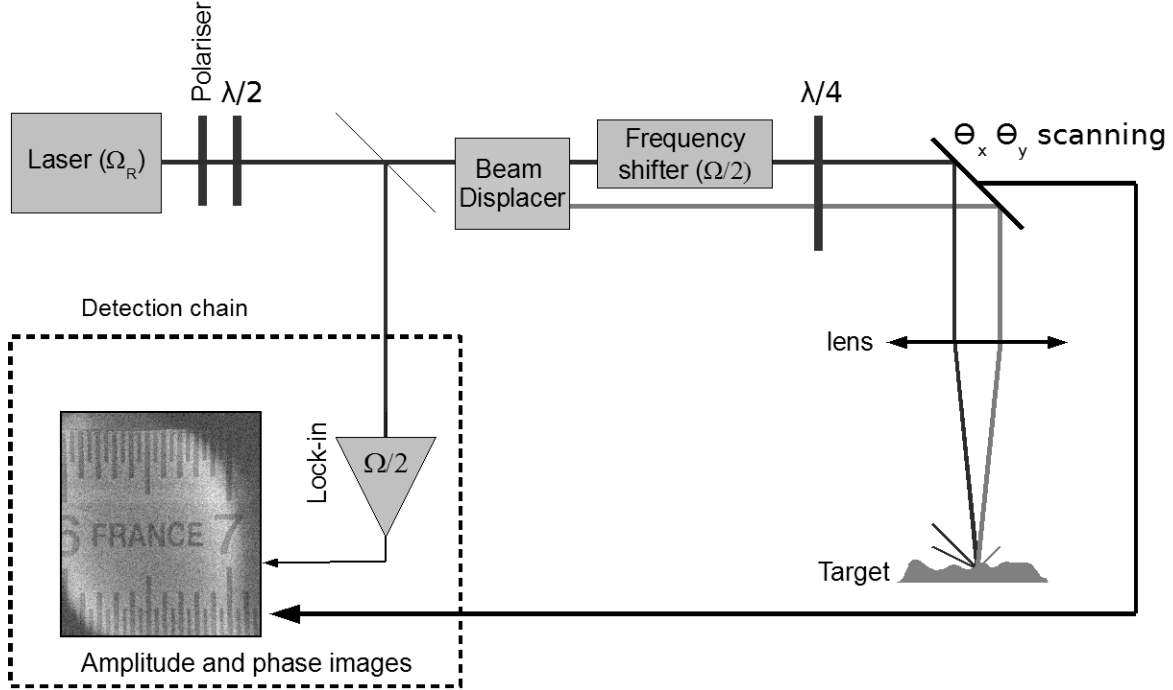

*Figure 3: quasi self aligned experimental setup insensitive to optical parasitic reflection.*

To optically isolate the parasitic reflection, we control the path of the backscattered light, in order to have different path the parasitic backscattering and the target backscattering. The selecting backscattered light path is obtained using birefringent properties of the beam displacer where the optical paths depend on the entering laser beam polarization. This used principle is illustrated in Fig.4. The beam displacer is Calcite crystal plate with a optical axis inclined to crystal plate edges. At the entrance of the beam displacer, we may decompose the laser beam polarization into two orthogonal linear polarized components, one polarized perpendicularly to the optical axis, called ordinary (o) component (corresponding to ordinary ray), and one with its polarization in a plan which includes the optical axis, called the extraordinary (e) component (corresponding to extraordinary ray). In Fig.4, the ordinary ray is not displaced and the extraordinary ray exits at a distance away from the incident beam. In Fig.4a, we consider an incident light with ordinary polarization and a parasitic reflection situated beyond the quarter-wavelength retardation plate which is oriented at the 45° angle to (e) and (o) polarization directions. After the reflection, the entering light in the beam splitter has an extraordinary polarization because the incident light has crossed twice the quater-wavelength retardation plate which rotates initial polarization by 90°. Consequently, the reflected laser beam is displaced relative to incident beam and it cannot be reinjected into laser cavity. We can do a similar reasoning for an incident extraordinary ray (fig.4b). For the backscattered light by the target, the light has two possibilities to come back toward the laser (identical path for incident and backscattered light or loop path). For the identical paths case, the light cannot be reinjected into the laser, similarly to the cases illustrated in fig.4a) and 4b). For the loop path case, the backscattered light by the target may be reinjected into the laser due to 90° polarization rotation. The ordinary incident ray comes back toward the laser by the extraordinary polarization path and vice versa (fig.4c).

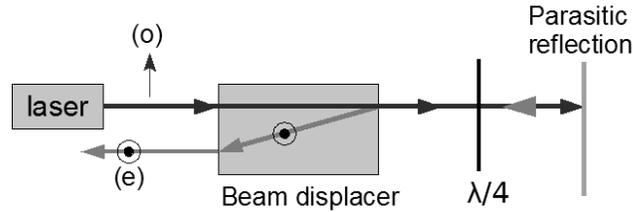

*fig.4 a)*

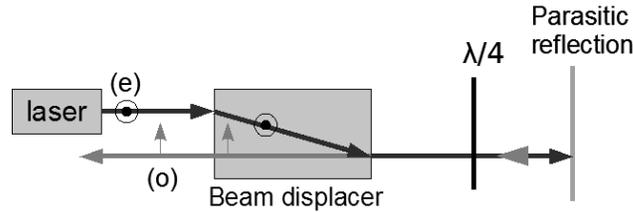

*fig.4 b)*

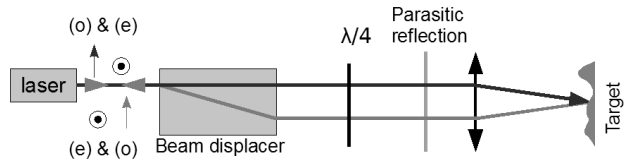

*fig.4 c)*

*Figure 4: principle of the optical isolation. (o)ordinary polarization and (e) extraordinary polarization.*

The electronic isolation is obtained by selecting the reference frequency of the lock in amplifier as in the setup proposed in [7]. Indeed, for a parasitic reflection, the laser/target and the target/ laser paths are necessary the same which means that the frequency shift is Ω if the optical path is the bottom arm in Fig.3 or zero if the optical path is the top arms in Fig.3. On the other hand, the frequency shift is Ω/2 for backscattered light by the target because the laser/target and the target/ laser path are necessary different. A detection at the Ω/2 frequency allows to avoid the adverse effect of parasitic reflection.

This double isolation ensures phase end amplitude image without the parasitic reflection contribution for any position of parasitic reflection in the setup. Indeed, if the parasitic reflection source is situated before the quarter-wavelength retardation plate then the isolation is electronic. If it is beyond the quarter-wavelength retardation plate then the isolation is both electronic and optical. As a result, optics used in the setup presented in Fig.3 doesn't need AR coating at the working wavelength, despite the extreme sensitivity of the LOFI technique. If the target completely depolarized the incident light then the quantity of reinjected light into the laser is devised by two, but the parasitic reflection is not affected if the parasitic reflection source do not depolarize the incident light.

## Experimental Results

In order to validate the principle of the optical device presented in figure 3 we have realized the same experiment as the one described in section II. The parasitic optical feedback is generated by a microscope slide placed on both arms beyond the quarter-wavelength retardation plate. We adjust the slide orientation in order to obtain $r_P \approx r_t$. We have realized phase and amplitude images at the frequency $\Omega/2$. The results obtained are given in Fig.5. Despite a significant parasitic reflection, we obtain phase and amplitude images of comparable quality as the ones obtained without parasitic reflection in section II (fig.2a, fig.2b). These results demonstrate that the adverse effects caused by parasitic reflections can be eliminate with the optical system proposed in figure 3.

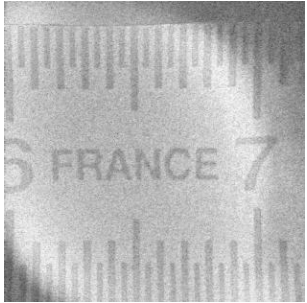
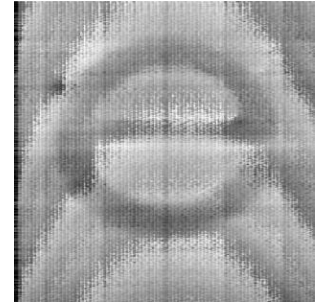

*fig5. a)*  *fig5. b)*

*figure 5: images obtained with parasitic reflection located beyond the quarter wavelength retardation plate in proposed quasi self aligned setup. a) Amplitude image realized at the frequency $\Omega/2$. b) Phase image realized at the frequency $\Omega/2$.*

In Figure 2a, the image is slightly truncated, we assume that this effect can be generated by overlapping variations between both beams on the target during the scanning, or by polarization variation of reinjected light through the polarizer during the scanning. During the scanning the incident angle of the ordinary and extraordinary laser beams changes and this might introduce a different reflectivity for both polarizations. Consequently, the polarization of the reinjected light through the polarizer may be different from the polarization incident light. This difference depends on the scanner angle meaning that the reinjected light into the laser might decrease for high angles. The use of a beam displacing prism with a smaller displacement should reduce this scanning effect.

## Conclusion

In this paper, we have proposed a self aligned setup for LOFI technique insensitive to parasitic optical feedback. Experimental results demonstrate that parasitic reflection contribution in phase and amplitude image can be eliminate while keeping the self aligned feature of the LOFI setup. Moreover, the proposed device allows a double isolation of parasitic reflections. The isolation may be electronic and /or optics depending on the location of the parasitic reflection. This double isolation allows using no AR coated optics in the setup despites the extreme sensitivity to optical feedback of the LOFI technique.  A successful experimental phase and amplitude images obtained with significant parasitic reflection validates the insensitivity to parasitic optical feedback. In future work, we plan to decrease the distance between both laser beams to limit truncation effect, and implement to LOFI microscopy setup the principle presented in this paper.